# Inducement of superconductivity in Fe(Te,S) by sulfuric acid treatment

Masanori NAGAO[*], Yoshikazu MIZUGUCHI[1], Keita DEGUCHI[2], Satoshi WATAUCHI, Isao TANAKA and Yoshihiko TAKANO[2]

*University of Yamanashi, 7-32 Miyamae, Kofu, 400-8511*

[1]*Tokyo Metropolitan University, Hachioji, Tokyo 192-0397*

[2]*National Institute for Materials Science, 1-2-1 Sengen, Tsukuba, 305-0047*

*Corresponding Author

Masanori Nagao

E-mail address: mnagao@yamanashi.ac.jp

Postal address: University of Yamanashi, Center for Crystal Science and Technology

Miyamae 7-32, Kofu 400-8511, Japan

Telephone number: (+81)55-220-8610

Fax number: (+81)55-254-3035

Since the discovery of Fe-based superconductors, much effort has been focused on search for new superconductivity in related compounds[1]. Fe chalcogenides are the simplest Fe-based superconductors[2]. S-substituted FeTe (FeTe$_{1-x}$S$_x$) is one of the Fe-chalcogenide superconductors with transition temperature $T_c \sim 10$ K. However, FeTe$_{1-x}$S$_x$ synthesized using a conventional solid-state reaction method does not show superconductivity[3]. The superconductivity is induced in FeTe$_{1-x}$S$_x$ by air exposure[4], oxygen annealing[5,6], and immersion in water[7] and alcoholic beverages[8]. Recently, Deguchi et al. reported that superconductivity of FeTe$_{1-x}$S$_x$ was induced by two ways. One way is oxidization of excess Fe, which exists at the interlayer site of FeTe$_{1-x}$S$_x$ structure and interferes with the appearance of superconductivity in FeTe$_{1-x}$S$_x$, by air exposure and oxygen annealing. Another way is the removal of excess Fe from the interlayer site of FeTe$_{1-x}$S$_x$ structure by the alcoholic treatment. If the excess Fe in FeTe$_{1-x}$S$_x$ structure could be removed efficiently, superconductivity of FeTe$_{1-x}$S$_x$ should be dramatically improved[9]. In this paper, we report the inducement and enhancement of superconductivity in Fe$_{0.95}$Te$_{0.85}$S$_{0.15}$ by the sulfuric acid treatment.

The polycrystalline samples of Fe$_{0.95}$Te$_{0.85}$S$_{0.15}$ were prepared using the solid-state reaction method. Fe (99.9%) powder, Te (99%) grains and S (99% up) powder were mixed in a nominal composition of Fe$_{0.95}$Te$_{0.85}$S$_{0.15}$ using a mortar, and sealed into a quartz tube in vacuum. And then the sample was heated at 700 °C for 20 h, and furnace-cooled. The part of the as-grown sample was characterized immediately by magnetic susceptibility measurement.

The remained powder was immersed into a diluted sulfuric acid solution in a glass bottle at room temperature for 5 and 50 h. Some samples were heated at 50-100 °C for 5 h. For comparison, the reference samples were prepared by immersing in pure water at room temperature. The treated samples were taken out from the bottle using filtration, dried in an oven at 70 °C for 10 min, and used immediately for measurement of the superconducting properties. Temperature dependence of the magnetic susceptibility in zero-field cooling (ZFC) condition was measured down to 5 K under a magnetic field of 10 Oe using a superconducting quantum interference device (SQUID) magnetometer. The Fe concentration in sulfuric acid solution after the treatment was analyzed by an inductively coupled plasma (ICP) spectroscopy. The samples were identified by powder X-ray diffraction (XRD) method with $CuK_\alpha$ radiation. We confirmed that there was almost no difference in the X-ray diffraction pattern between the as-grown and the treated samples within the sensitivity of lab-level x-ray powder diffraction.

Figure 1 shows the temperature dependence of normalized susceptibility for the three samples in pure water and diluted sulfuric acid solution (10 and 47 wt%) treatment at room temperature for various periods. For samples treated in pure water, weak and broad superconducting transition was observed at around 8 K. In contrast, the superconducting transitions of the samples in diluted sulfuric acid solution treatment were sharper than that in pure water. Superconducting signal was increased with increasing treatment period. These results suggest that the sulfuric acid solution treatment induces superconductivity in $Fe_{0.95}Te_{0.85}S_{0.15}$ as compared to that in pure water treatment.

According to the previous research[7], increasing the reacting temperature of immersed samples enhances reaction speed. Figure 2 shows the temperature dependence of normalized susceptibility for the $Fe_{0.95}Te_{0.85}S_{0.15}$ samples in 10 wt% diluted sulfuric acid solution treatment at various heating temperatures for 5 h. Superconducting signal increased with increasing temperature up to 80 °C. However, the signal for 100 °C was smaller than that for 80 °C.

We measured mass fraction of the dissolved Fe from the $Fe_{0.95}Te_{0.85}S_{0.15}$ samples treated by the sulfuric acid solution, and examined that dependence of normalized susceptibility on the mass fraction of the dissolved Fe for the treated $Fe_{0.95}Te_{0.85}S_{0.15}$ samples shown in Fig. 3. In the case of treatment in 47 wt% sulfuric acid solution of at 80 °C, the superconducting signal (absolute value of normalized susceptibility) was smaller than that in 10 wt% sulfuric acid solution. In contrast, the superconducting signal increased with the increase in the mass fraction of the dissolved Fe. The superconductivity in $Fe_{0.95}Te_{0.85}S_{0.15}$ is related to mass fraction of the dissolved Fe. This result is considered to be due to deintercalation of the excess Fe at the interlayer in $Fe_{0.95}Te_{0.85}S_{0.15}$ sample.

In conclusion, we reported that sulfuric acid solution treatment induced superconductivity in $Fe_{0.95}Te_{0.85}S_{0.15}$. Sulfuric acid solution treatment is more effective as compared to pure water. It was found that the two post treatments, water and sulfuric acid immersions, can induce superconductivity, while the mechanism of the inducement of superconductivity is essentially different.

**Acknowledgment**

The authors would like to thank Dr. A. Miura of University of Yamanashi for useful discussions.

Fang, L. Spinu, P. Schiffer, Y. Liu, and Z. Q. Mao: Phys. Rev. B **80** (2009) 174509.

**Figure captions**

Figure 1 Temperature dependence of normalized susceptibility for the $Fe_{0.95}Te_{0.85}S_{0.15}$ samples treated at room temperature for various period in; (a) pure water, (b) 10 wt% diluted sulfuric acid, (c) 47 wt% diluted sulfuric acid.

Figure 2 Temperature dependence of normalized susceptibility for the $Fe_{0.95}Te_{0.85}S_{0.15}$ samples in 10 wt% diluted sulfuric acid treatment at various heating temperature for 5 h.

Figure 3 Relationship between mass fraction of dissolved Fe and absolute value of normalized susceptibility at 5 K for the $Fe_{0.95}Te_{0.85}S_{0.15}$ samples in various treatment conditions.

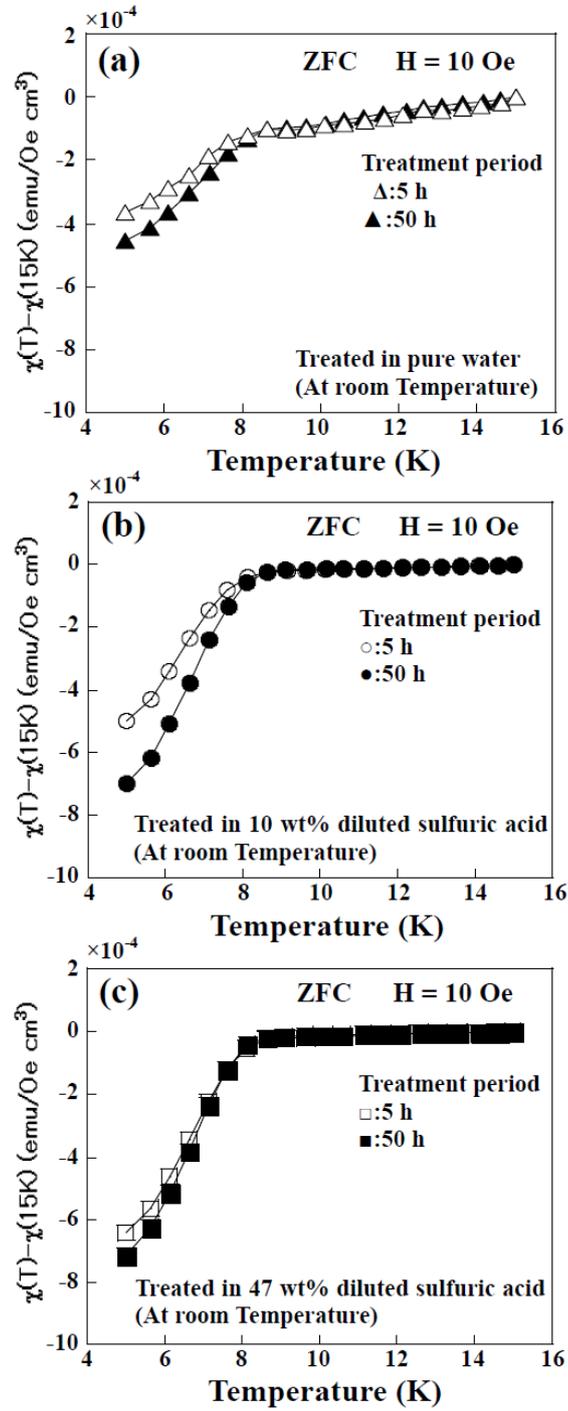

**Figure 1**

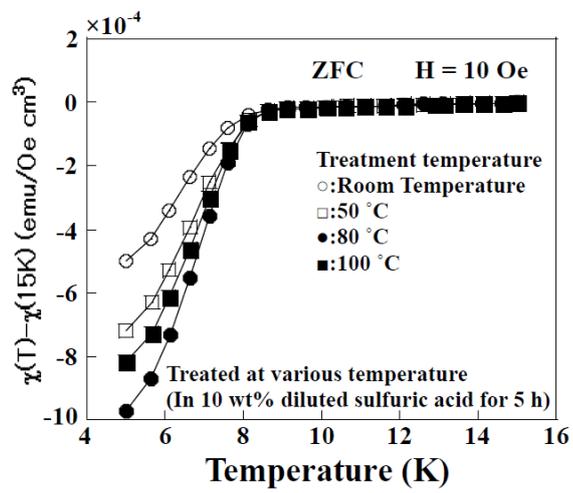

**Figure 2**

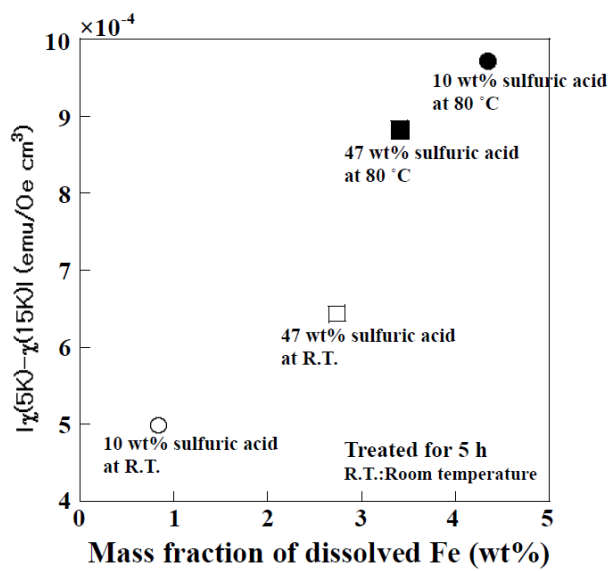

**Figure 3**